# A new algebraic approach to the graph isomorphism and clique problems

Roman Galay

Daniil Kalistratov

As it follows from Gödel's incompleteness theorems, any consistent formal system of axioms and rules of inference should imply a true unprovable statement. Actually this fundamental principle can be efficiently applicable in computational mathematics and complexity theory concerning the computational complexity of problems from the class NP, particularly and especially the NP-complete ones. While there is a wide set of algorithms for these problems that we call heuristic, the correctness or/and complexity of each concrete algorithm (or the probability of its correct and polynomial-time work) on a class of instances is often too difficult to determine, although we may also assume the existence of a variety of algorithms for NP-complete problems that are both correct and polynomial-time on all the instances from a given class (where the given problem remains NP-complete), but whose correctness or/and polynomial-time complexity on the class is impossible to prove as an example for Gödel's theorems. However, supposedly such algorithms should possess a certain complicatedness of processing the input data and treat it in a certain algebraically "entangled" manner. The same algorithmic analysis in fact concerns all the other significant problems and subclasses of NP, such as the graph isomorphism problem and its associated complexity class GI.

The following short article offers a couple of algebraically entangled polynomial-time algorithms for the graph isomorphism and clique problems whose correctness is yet to be determined either empirically or through attempting to find proofs.

Besides, the paper contains a description of an equation system for elements of a set of groups (which can also be interpreted as an algebraic equation system) that can be polynomial-time reduced to a graph isomorphism problem and, in the same time, is a non-linear extension of a system of modular linear equations where each equation has its own modulus (hence implying the question whether it's NP-complete).

The authors are grateful to Prof. Anuj Dawar (Professor of Logic and Algorithms in the Department of Computer Science and Technology, Robinson College, University of Cambridge) for kindly endorsing the present article for publishing in arXiv.

# An heuristic polynomial-time algorithm for the graph isomorphism problem.

**Notation:**

For a real-valued matrix B, by Sp(B) let's denote the set of values its entries are equal to, and by MSp(B) the multi-set of those values (including the multiplicity of each value). We'll call those set and multi-set the *entry spectrum* and the *entry multi-spectrum* of B correspondingly.

Given two simple undirected graphs $G_1$ and $G_2$ with n vertices whose adjacency matrices are $A_1$ and $A_2$ correspondingly, we're going to build two sequences of real-valued matrices $A_1^{(i)}$ and $A_2^{(i)}$ by the following recursive scheme:

$$A_1^{(0)} = A_1, \qquad A_2^{(0)} = A_2$$

Beginning with i = 1, at the step i of our recursive process, first of all we determine whether $MSp(A_1^{(i)}) = MSp(A_2^{(i)})$.

If it's not so then $G_1$ and $G_2$ are definitely non-isomorphic. If, otherwise, the equality holds then we create a random $|Sp(A_1^{(i)})|$-vector y and in both matrices $A_1^{(i)}$ and $A_2^{(i)}$ we replace each entry whose value is the j-th element of $Sp(A_1^{(i)})$ by $y_j$. After that, we choose a random polynomial p(t) of degree n-1 and calculate $A_1^{(i+1)} = p(A_1^{(i)})$ and $A_2^{(i+1)} = p(A_2^{(i)})$.

We stop the whole process when either the entry multi-spectrums of the two current matrices are different or the cardinality of their common entry spectrum doesn't increase any more from one step to another, i.e. $|Sp(A_1^{(k)})| = |Sp(A_1^{(k-1)})|$ (accordingly, the overall number k of performed steps can't exceed $n^2$).

We hence declare the initial graphs' non-isomorphism in the former case, and their isomorphism in the latter one.

When the process is stopped, with the overall number of steps equal to k and the final entry spectrum $\{\alpha_1, \ldots, \alpha_m\}$, there should exist symmetric 0,1-matrices $H_1^{(1)}, \ldots, H_1^{(m)}, H_2^{(1)}, \ldots, H_2^{(m)}$ such that

$$A_1^{(k)} = \sum_{u=1}^{m} \alpha_u H_1^{(u)}, \qquad A_2^{(k)} = \sum_{u=1}^{m} \alpha_u H_2^{(u)},$$

We can consider, for u = 1,..,m, $H_1^{(u)}$ and $H_2^{(u)}$ as the adjacency matrices of some graphs $G_1^{(u)}$ and $G_2^{(u)}$. In case if the initial graphs $G_1$ and $G_2$ are really isomorphic, those two graphs should be isomorphic as well. Moreover, due to the entry spectrum's non-growing at the end of the above-described process, for any pair $v, w \in \{1, \ldots, m\}, v \neq w$, $G_1^{(v)}$ and $G_1^{(w)}$ should possess no common edges and $G_2^{(v)}$ and $G_2^{(w)}$ should too; also there should exist a sequence of coefficients

$d_1^{(v,w)}, \ldots, d_m^{(v,w)}$ such that $H_1^{(v)}H_1^{(w)} + H_1^{(w)}H_1^{(v)} = \sum_{u=1}^{m} d_u^{(v,w)} H_1^{(u)}$ and $H_2^{(v)}H_2^{(w)} + H_2^{(w)}H_2^{(v)} = \sum_{u=1}^{m} d_u^{(v,w)} H_2^{(u)}$ (the latter condition also concerns the case v = w). Verifying these additional relations is the proposed algorithm's final action when declaring the isomorphism of $G_1$ and $G_2$. Thus we get a pair of identical commutative algebras that are two linear spaces with the bases $H_1^{(1)}, \ldots, H_1^{(m)}$ and $H_2^{(1)}, \ldots, H_2^{(m)}$ whose elements are $X = \sum_{u=1}^{m} x_u H_1^{(u)}$ and $X = \sum_{u=1}^{m} x_u H_2^{(u)}$ correspondingly and whose algebra product is defined as $X * Y = XY + YX$.

However, if, after some global steps, in each of the above linear spaces we take "standard" matrix products of various space elements (related by having the same coefficients for each space) then we'll receive, upon the entry-spectrum's ceasing to grow with further performed steps, a pair of corresponding identical non-commutative algebras whose algebra product is just the standard matrix product (instead of the above-introduced symmetric one) and whose product closure condition is, accordingly, $H_1^{(v)}H_1^{(w)} = \sum_{u=1}^{m} d_u^{(v,w)} H_1^{(u)}$ and $H_2^{(v)}H_2^{(w)} = \sum_{u=1}^{m} d_u^{(v,w)} H_2^{(u)}$ for any $v, w \in \{1, \ldots, m\}$ (instead of the above symmetric one).

The above-formulated algorithm can be naturally adjusted for digraphs with absolutely the same computational circuit (while, though strange, generating, in the commutative algebra case, commutative algebras at the very end as well and, in case if the chosen field is of characteristic 2, we'll receive Lie algebras). Moreover, it can be generalized for an arbitrary field (instead of $\mathbb{R}$) with all the computations performed over that field.

Another type of generalization this algorithm might be subjected to is as follows.

**Definition:**

Given an n×n- matrix B, let's define a product $B^{r_1} J_n B^{r_2} J_n \ldots B^{r_{s-1}} J_n B^{r_s}$ (where $J_n$ is an n×n-matrix all whose entries equal unity and $r_1, \ldots, r_s$ are non-negative integers unexceeding n-1) as its *meta-power* of *meta-degree* $(r_1, r_2, \ldots, r_s)$ and a linear combination of a set of its meta-powers with coefficients taken from a chosen field as a *meta-polynomial* in B over the field.

If an n×n-matrix can be turned into another one via permuting its rows and columns by a permutation π (i.e. if we have a pair of isomorphic matrices) then an arbitrary meta-polynomial computed in both of them should give us a pair of isomorphic matrices as well (with the same transitional permutation π). Hence the idea of replacing, in the proposed algorithmic approach, random polynomials p(t) of degree n-1 by random meta-polynomials may look yet perspective, even though it's still difficult to figure out how the meta-polynomials' sets of utilized meta-degrees could be restricted in such a case. Nevertheless, the principle of stopping the recursive process upon getting either different entry multi-spectrums of the two current matrices or their common entry spectrum's cardinality ceasing to grow from one step to another remains intact (while the matrices $A_1^{(i)}$ and $A_2^{(i)}$ generically cease to be symmetric for i > 0 even in the case of undirected initial graphs). The final splittings of the two matrices $A_1^{(k)}$ and $A_2^{(k)}$ (where k is the overall number of performed steps) apparently will have nearly the same structure as in the case of random polynomials p(t), but with the only additional condition of $H_1^{(u)} J_n$ and $J_n H_1^{(u)}$ belonging to the first final algebra and $H_2^{(u)} J_n$ and $J_n H_2^{(u)}$ to the second one for u = 1,…,m.

In case if we deal with a problem of partitioned graph isomorphism where the set of vertices V is partitioned into subsets $V_1, \ldots, V_m$ and the task is to determine whether there exists an isomorphism from $G_1$ to $G_2$ mapping $V_k$ to itself for k = 1,…,m, the matrix $J_n$ can be replaced by the block-diagonal matrix $J(V_1, \ldots, V_m)$ whose k-th diagonal block is $J_{|V_k|}$ for k = 1,…,m.

Such a final algebra doesn't depend on the chosen random polynomials or, generally, meta-polynomials within the algorithmic circuit (as well as the random substitution vectors we use for entry spectrums' replacements). A proof of this fact can be received via using, instead of random elements of the basic field, independent indeterminants as the entries of substitution vectors in each spectrum replacement and the coefficients of utilized polynomials at each global step and getting, accordingly, polynomials in those indeterminants as the entries of transformed matrices (in such a case the algorithm ceases to be polynomial-time, though). We'll call it a *splitting algebra* of a graph that is a system of its invariants (though, depending on the context, sometimes we'll also interpret it as the corresponding set of matrices $\sum_{u=1}^{m} x_u H_1^{(u)}$ which isn't invariant), while an associative splitting algebra will be also called a *splitting ring*.

**Definition:**

Given a set of n×n-matrices S over F and an algorithm $\mathfrak{A}$ for receiving a splitting algebra over a field F (we'll call it the *algebra builder* that is supposed to be defined uniformly for all n), by Alg(S, $\mathfrak{A}$) we'll denote the splitting algebra received via applying $\mathfrak{A}$ to a formal (i.e. with independent formal variables taken as coefficients) linear combination of S over F.

As it was stated earlier, Alg(S,$\mathfrak{A}$) can be received via taking a random set of values of the formal coefficients. When the generating set S isn't given, we'll just write Alg(…,$\mathfrak{A}$) what will be equivalent to referring to the algebra builder $\mathfrak{A}$. Also, in the partial case of a singleton S containing just one matrix A, we'll write Alg(A,$\mathfrak{A}$).

Hence we can reduce, via the proposed method, determining whether two graphs are isomorphic to determining whether a related (by having identical sequences of coefficients at $H_1^{(u)}$ and $H_2^{(u)}$) random pair of elements of their splitting algebras represented by the corresponding matrices $\sum_{u=1}^{m} x_u H_1^{(u)}$ and $\sum_{u=1}^{m} x_u H_2^{(u)}$ is isomorphic as a pair of weighted graphs (or digraphs), hence providing new splitting options for possibly building some extensions of the splitting algebras.

In this regard, it would be worth noting that all the above-described algebras can't separate two non-isomorphic strongly regular graphs with the same set of parameters (as the square of a strongly regular n-graph's adjacency matrix is a linear combination of its adjacency matrix, $I_n$ and $J_n$), even though they do work out for nearly all graphs. Therefore some more refined transformations preserving the set of isomorphisms between two matrices are needed for to resolve the hardest cases of graph isomorphism.

**Definition:**

Given an r×r-matrix invariant I and an r×r-matrix algebra with the basis $H^{(1)}, \ldots, H^{(m)}$ over a basic field, we'll call $\mathrm{Alg}_I\big(A^{(\{1,\ldots,n\}\backslash L, \{1,\ldots,n\}\backslash L)}, \mathfrak{A}\big) = \sum_{t=1}^{m} I(H^{(t)}) H^{(t)}$ its I-element.

**Definition:**

Given a splitting algebra $\mathrm{Alg}(\ldots, \mathfrak{A})$ and an (n-k)×(n-k)-matrix invariant I, by its k-th I-derivative $\mathrm{Alg}_I^{(k)}(\ldots, \mathfrak{A})$ (or its k-th derivative in I) we'll understand the sum, over all the subsets $L \subseteq \{1, \ldots, n\}$ of cardinality k, of the matrices received via replacing all the entries lying in either rows or columns from the set L by zero and the principal submatrix $A^{(\{1,\ldots,n\}\backslash L, \{1,\ldots,n\}\backslash L)}$ by $\mathrm{Alg}_I\big(A^{(\{1,\ldots,n\}\backslash L, \{1,\ldots,n\}\backslash L)}, \mathfrak{A}\big)$.

**Definition:**

for a real m×n-matrix B of rank m, m ≤ n, let's define its j-th LP-height $\mathrm{height}_j(B)$ as the optimal goal value of the LP-problem

$$Bx \leq B\vec{1}_n, \quad x \in \mathbb{R}^n,$$

$$x_i \geq 0, \quad i = 1, \ldots, n$$

$$x_j \to \max$$

and the vector $\mathrm{height}(B) = \{\mathrm{height}_j(B)\}_n \in \mathbb{R}^n$ as its LP-height vector (or just its height-vector).

**Definition:**

For a real symmetric n×n-matrix $M = PJP^{-1} = \sum_{k=1}^{|\lambda|} \lambda_k P_k P_k^T$ where J is its diagonal Jordan form, P is an orthogonal matrix (i.e. its orthonormal Jordan basis), λ is its eigenvalue spectrum and the matrix $P_k$ is composed of its eigenvectors corresponding to the eigenvalue $\lambda_k$ for k = 1,…,|λ|, we'll define $\mathrm{height}(P_k^T)$ as M's eigenheight-vector corresponding to the eigenvalue $\lambda_k$ and, generally, $\mathrm{height}\begin{pmatrix} P_{k_1}^T \\ \ldots \\ P_{k_q}^T \end{pmatrix}$ as M's eigenheight-vector corresponding to the subset $\{\lambda_{k_1}, \ldots, \lambda_{k_q}\}$ of M's eigenvalue spectrum.

Hence adding the matrix $\mathrm{height}\begin{pmatrix} P_{k_1}^T \\ \ldots \\ P_{k_q}^T \end{pmatrix} \mathrm{height}^T \begin{pmatrix} P_{k_1}^T \\ \ldots \\ P_{k_q}^T \end{pmatrix}$ multiplied by an arbitrary scalar to the matrix M won't change its automorphism group and, therefore, subjecting two matrices with the same eigenvalue spectrum to such a transformation corresponding to the same subset of their common spectrum won't change the set of isomorphisms between them.

Altogether, we hence got one more isomorphism-preserving transformation that can be applied, along with the earlier introduced meta-polynomial transformations, for searching a difference between two matrices in case if they're not isomorphic, while each new transformation generates an extension of their splitting algebras. Besides, as it was also introduced above, any splitting algebra yields its derivatives, and those derivatives of a bounded degree preserve a polynomial-time computable algebra's belonging to the complexity class P what provides us with another family of polynomial-time computable extensions of any splitting algebra.

It would be also worth noting that other types of splitting algebras' extensions might be based on other compatible (with any isomorphism) matrix decompositions such as the polar decomposition. In this regard, further we're going to formalize this principle, as well to give some set-theoretical basis to all the above-stated algorithmic schemes.

**Notation:**

Given a permutation $\pi \in S_n$, by $I_\pi$ we'll denote the matrix received from $I_n$ via subjecting its rows to the permutation $\pi$.

**Definition:**

Given a ring K, a transformation $\omega: K^{n \times n} \to K^{n \times n}$ (applicable for any natural n) will be called isomorphism-commuting if $\forall \pi \in S_n, \forall X \in K^{n \times n}$  $\omega(I_\pi X I_\pi^T) = I_\pi \omega(X) I_\pi^T$.

**Lemma 1:**

The function composition of two isomorphism-commuting transformations is an isomorphism-commuting transformation, i.e. the set of isomorphism-commuting transformations is a semigroup under the function composition operation.

**Lemma 2:**

Given a square matrix A over a ring K, the set of its isomorphism-commuting transformations is a ring (that is an extension of K) under the matrix arithmetic operations.

Hence the set of isomorphism-commuting transformations has the two important properties that it's closed under function composition and the matrix arithmetic operations. However, the same two properties are possessed by the set of isomorphism-commuting transformations computable in polynomial time. Let's denote the two sets by ICT and ICTP correspondingly.

**Definition:**

Given a ring K, a function $f: K^{n \times n} \times \ldots \times K^{n \times n} \to K^{n \times n}$ in m n×n-matrix variables $X_1, \ldots, X_m$ (applicable for any natural n) will be called isomorphism-commuting if $\forall \pi \in S_n, \forall X_1, \ldots, X_m \in K$  $f(I_\pi X_1 I_\pi^T, \ldots, I_\pi X_m I_\pi^T) = I_\pi f(X_1, \ldots, X_m) I_\pi^T$.

We hence can consider an isomorphism-commuting transformation as an isomorphism-commuting function in one variable. Besides, the set of isomorphism-commuting functions is also closed under function composition and the matrix arithmetic operations.

**Definition:**

Given a set $f = \{f_0(X_1, \ldots, X_m), f_1(X_1, \ldots, X_m), \ldots, f_r(X_1, \ldots, X_m)\}$ of r+1 n×n-matrix functions in m n×n-matrices and an n×n-matrix A over a ring K such that the matrix equation system

$$f_0(X_1, \ldots, X_m) = Y$$

$$f_q(X_1, \ldots, X_m) = 0_{n \times n} \quad \text{for q = 1,...,r}$$

has a unique solution $\text{Decom}_1(Y, f), \ldots, \text{Decom}_m(Y, f)$ for any $Y \in K^{n \times n}$. Then $\text{Decom}_1(A, f), \ldots, \text{Decom}_m(A, f)$ will be called the f-decomposition of A over K.

**Lemma 3:**

Given a set $f = \{f_0(X_1, \ldots, X_m), f_1(X_1, \ldots, X_m), \ldots, f_r(X_1, \ldots, X_m)\}$ of isomorphism-commuting functions over a ring K, $\text{Decom}_1(A, f), \ldots, \text{Decom}_m(A, f)$ are isomorphism-commuting transformations of a square matrix A over K.

**Conjecture 1:**

For any ring K, any two square matrices A, B over K are isomorphic if and only if $\text{MSp}(\omega(A)) = \text{MSp}(\omega(B))$ for any isomorphism-commuting transformation.

**Conjecture 2:**

For any ring K, there exists a polynomial p(n) such that, for any two n×n-matrices A, B over K, A and B are isomorphic if and only if $\text{MSp}(\omega(A)) = \text{MSp}(\omega(B))$ for any isomorphism-commuting transformation computable in p(n) arithmetic operations over K.

If the latter conjecture is true then the question whether two given matrices are isomorphic can be answered, by an algorithm from the complexity class RP, via subjecting both matrices to a random isomorphism-commuting transformation computable in p(n) arithmetic operations over the basic ring (i.e. a random element of ICTP generating, accordingly, the algorithm's polynomial-time complexity) and determining whether the entry multi-spectrums of the two transformed matrices are equal. In such a case we'll also get a randomized complete system of graph invariants, although, as it was already mentioned earlier for splitting algebras received via the use of eigenvectors, there might exist a fixed set of ITCP's elements that is the complete invariant system.

**A special case of the graph isomorphism problem that is a system of group equations.**

Let's consider a special case of the isomorphism problem for two colored graphs (where all the vertices and edges are given colors) $G_1 = (V, E_1)$ and $G_2 = (V, E_2)$ such that

$$V = \cup_{i=1}^{m} V^{[i]}, \ \forall v \in V^{[i]} \ \text{color}(v) = c_i,$$

$$V^{[j]} \cap V^{[k]} = \emptyset \ \text{and} \ c_j \neq c_k \ \text{if} \ j \neq k$$

In such a case we get the following system of equations:

$$\psi_k \psi_j G(V^{[j]} \cup V^{[k]}, E_1(V^{[j]}, V^{[k]})) = G(V^{[j]} \cup V^{[k]}, E_2(V^{[j]}, V^{[k]}))$$

where $G\left(V^{[j]} \cup V^{[k]}, E_1(V^{[j]}, V^{[k]})\right), G(V^{[j]} \cup V^{[k]}, E_2(V^{[j]}, V^{[k]}))$ are the bipartite graphs generated by $G_1, G_2$ correspondingly on the set $V^{[j]} \cup V^{[k]}$ as the induced graphs where all the internal edges inside the sets $V^{[j]}, V^{[k]}$ were eliminated, and $\psi_i$ is a permutation of $V^{[i]}$ (considered as a transformation of any graph whose vertex set contains $V^{[i]}$) belonging to the set of isomorphisms between $G(V^{[i]}, E_1(V^{[i]}))$ and $G(V^{[i]}, E_2(V^{[i]}))$ where $G(V^{[i]}, E_1(V^{[i]}))$, $G(V^{[i]}, E_2(V^{[i]}))$ are the graphs induced by $V^{[i]}$ in $G_1, G_2$ correspondingly. As we can represent any $\psi_i$ as $\rho_i \varphi_i$ where $\rho_i$ is a fixed isomorphism between $G(V^{[i]}, E_1(V^{[i]}))$ and $G(V^{[i]}, E_2(V^{[i]}))$ and $\varphi_i \in \mathrm{Aut}(G(V^{[i]}, E_1(V^{[i]})))$, we'll hence receive the system of group equations for $\varphi_1, \ldots, \varphi_m$

$$\rho_j \varphi_j \rho_k \varphi_k G(V^{[j]} \cup V^{[k]}, E_1(V^{[j]}, V^{[k]})) = G(V^{[j]} \cup V^{[k]}, E_2(V^{[j]}, V^{[k]})) \quad j, k = 1, \ldots, m, j \neq k$$

The latter equation for a given pair j, k is a requirement for $\rho_j \varphi_j$, $\rho_k \varphi_k$ to transform the $|V^{[j]}| \times |V^{[k]}|$-matrix $M_1^{[j,k]}$ generated by $G(V^{[j]} \cup V^{[k]}, E_1(V^{[j]}, V^{[k]}))$ as the labeled (according to the given colors) bipartite adjacency matrix (let's call it a **bridge-matrix**) between its parts $V^{[j]}$, $V^{[k]}$ into the corresponding matrix $M_2^{[j,k]}$ generated by $G(V^{[j]} \cup V^{[k]}, E_2(V^{[j]}, V^{[k]}))$ – while $\rho_j \varphi_j$ acts as a permutation of $M_1^{[j,k]}$'s rows and $\rho_k \varphi_k$ as a permutation of $M_1^{[j,k]}$'s columns.

We can also consider the two compacted weighted graphs $\mathrm{Com}(G_1), \mathrm{Com}(G_2)$ with the common vertex set $\{V^{[1]}, \ldots, V^{[m]}\}$ where the weight-graph of an existing (i.e. having at least one initial edge between the sets $V^{[j]}, V^{[k]}$) edge $(V^{[j]}, V^{[k]})$ is defined, for $j \neq k$, as $G(V^{[j]} \cup V^{[k]}, E_1(V^{[j]}, V^{[k]}))$ for $\mathrm{Com}(G_1)$ and as $G(V^{[j]} \cup V^{[k]}, E_2(V^{[j]}, V^{[k]}))$ for $\mathrm{Com}(G_2)$, and the loop $(V^{[i]}, V^{[i]})$'s weight-graph (or the loop-graph of $V^{[i]}$) is defined as $G(V^{[i]}, E_1(V^{[i]}))$ for $\mathrm{Com}(G_1)$ and as $G(V^{[i]}, E_2(V^{[i]}))$ for $\mathrm{Com}(G_2)$.

One interesting partial case of such a construction is a couple of bipartite weighted graphs $\mathrm{Com}(G_1), \mathrm{Com}(G_2)$ where the two connected parts their vertex set $\{V^{[1]}, \ldots, V^{[m]}\}$ is partitioned into are interpreted as the set of "variables" and the set of "equations". Particularly, let's consider the following subcase. Let any pair of corresponding "variables" have equal loop-graphs in $\mathrm{Com}(G_1), \mathrm{Com}(G_2)$, and any pair of corresponding "equations" do too. Also let the bridge-matrix $M_1^{[j,k]}$ between the j-th "variable" and the k-th "equation" have $m_{j,k}$ columns and possess a non-singleton group $\mathrm{Aut}_{\mathrm{row}}(j, k)$ of permutations $\mu$ of its rows such that each $\mu$ is an automorphism of the j-th "variable's" loop-graph and generates a non-empty group $\mathrm{Aut}_{\mathrm{column}}(\mu, j, k)$ of permutations of $M_1^{[j,k]}$'s columns mapping the matrix $I_\mu M_1^{[j,k]}$ back to $M_1^{[j,k]}$, while $M_2^{[j,k]}$ is received from $M_1^{[j,k]}$ via permuting its columns by a permutation $\sigma_{j,k} \in S_{m_{j,k}}$. We'll also denote $\mathrm{Aut}_{\mathrm{column}}(j, k) = \cup_{\mu \in \mathrm{Aut}_{\mathrm{row}}(j,k)} \mathrm{Aut}_{\mathrm{column}}(\mu, j, k)$ that is, obviously, a subgroup of $S_{m_{j,k}}$. Also let, in both graphs $\mathrm{Com}(G_1)$ and $\mathrm{Com}(G_2)$, the k-th "equation's" (whose set of "variables" it connected with we'll denote by $\mathrm{Var}(k)$) loop-graph be a weighted bipartite graph one of whose parts is connected with "variables" from $\mathrm{Var}(k)$ and another one is a "free" part such that the bridge-matrix between the two parts is composed, in an arbitrary order of columns, of all the column-vectors $\begin{pmatrix} I_{\pi_{j_{1,k},k}} g_{j_{1,k},k} \\ \cdots \\ I_{\pi_{j_{|Var(k)|,k},k}} g_{j_{1,k},k} \end{pmatrix}$ – where Var(k) = $\{j_{1,k}, \ldots, j_{|Var(k)|,k}\}$ and, for t = 1,…, |Var(k)|, $g_{j_{t,k},k}$ is an $m_{j_{t,k},k}$-vector with pair-wise distinct entries, $\pi_{j_{t,k},k}$ is an $m_{j_{t,k},k}$-permutation from a group $H_{j_{t,k},k}$ intersecting (in $S_{m_{j_{t,k},k}}$) with the left coset $\sigma_{j_{t,k},k} \text{Aut}_{\text{column}}(j_{t,k}, k)$ – such that $(\pi_{j_{1,k},k}, \ldots, \pi_{j_{|Var(k)|,k},k}) \in H_k = \langle S \rangle$ where S⊆ $H_{j_{1,k},k} \times \ldots \times H_{j_{|Var(k)|,k},k}$ and $H_k$ is the group generated by S (we'll call it the "equation's" *formula group*). And let all the columns of $M_1^{[j_{t,k},k]}$ and $M_2^{[j_{t,k},k]}$ non-incident to the t-th vector-coordinate of $\begin{pmatrix} I_{\pi_{j_{1,k},k}} g_{j_{1,k},k} \\ \cdots \\ I_{\pi_{j_{|Var(k)|,k},k}} g_{j_{|Var(k)|,k},k} \end{pmatrix}$ be equal to a zero column, for t = 1,…,|Var(k)| (and, for the purpose of simplicity, let's further understand by $M_1^{[j_{t,k},k]}$ and $M_2^{[j_{t,k},k]}$ just the matrices composed of all the columns incident to that t-th vector-coordinate).

For a permutation $\pi^{(j,k)} \in \text{Aut}_{\text{column}}(j,k)$, let's define its row-permutation $M_1^{[j,k]}$-generator group $\text{Aut}_{\text{column}}^{-1}(\pi^{(j,k)}, j, k) = \{\mu: \mu \in \text{Aut}_{\text{row}}(j,k), \pi^{(j,k)} \in \text{Aut}_{\text{column}}(\mu, j, k)\}$ that is $\text{Aut}_{\text{row}}(j,k)$'s subgroup consisting of all its elements generating, particularly, $\pi^{(j,k)}$. Actually, in a set-theoretic notation, $\text{Aut}_{\text{column}}^{-1}(\text{Aut}_{\text{column}}(j,k), j, k) = \text{Aut}_{\text{row}}(j,k)$. We'll also denote by Eq(j) the set of "equations" the "variable" j is involved in.

Altogether, when we have r "equations" and q "variables", we receive the following system of equations for the permutation variables $\pi^{(j,k)}$:

$(\sigma_{j_{1,k},k} \pi^{(j_{1,k},k)}, \ldots, \sigma_{j_{|Var(k)|,k},k} \pi^{(j_{|Var(k)|,k},k)}) \in H_k$ for k = 1,…,r that we'll call the "equation" conditions

and $\bigcap_{k \in Eq(j)} \text{Aut}_{\text{column}}^{-1}(\pi^{(j,k)}, j, k) \neq \emptyset$ for j = 1,…,q that we'll call the "variable" conditions.

The above two sets of conditions also imply the set of conditions $\sigma_{j,k} \pi^{(j,k)} \in H_{j,k}$ for all the given "variable"-"equation" edges (j,k) that we'll call the edge conditions.

However, if in some "equations" from a set $K_{\text{linear}} \subseteq \{1, \ldots, r\}$ we take the partial case of their formula groups generated by the equality $\prod_{t=1}^{|Var(k)|} \chi_{j_{t,k},k}(\pi_{j_{t,k},k}) = 1$, where $\chi_{j_{t,k},k}$ is a multiplicative character of $H_{j_{t,k},k}$ over a field $F_k$, then we'll receive the following subsystem of equations for k ∈ $K_{\text{linear}}$ and the variables $\pi^{(j,k)}$ such that $\sigma_{j,k} \pi^{(j,k)} \in H_{j,k}$:

$\sum_{j \in Var(k)} \log_{a_k}(\chi_{j,k}(\pi^{(j,k)})) = -\sum_{j \in Var(k)} \log_{a_k} \chi_{j,k}(\sigma_{j,k})$ (mod (|$F_k$| - 1))

(where $a_k$ is a generating element of the field $F_k$).

If we also simplify all the bridge-matrices between the "variables" and the "equations" to identity matrices then, for any "variable"-"equation" edge (j,k), $\text{Aut}_{\text{row}}(j,k)$ would coincide with $\text{Aut}_{\text{column}}(j,k)$ and, for any j, the j-th "variable" condition would become the condition that for any k′, k″ ∈ Eq(j) $\pi^{(j,k')} = \pi^{(j,k'')} = \pi^{(j)}$ where $\pi^{(j)} \in \bigcap_{k \in Eq(j)} H_{j,k}$ is an automorphism of the j-th "variable's" loop-graph. Then the whole system of equations would turn into the set $K_{linear}$ of linear equations for the values $\log_{a_k}(\chi_{j,k}(\pi^{(j,k)})) = \log_{a_k}(\chi_{j,k}(\pi^{(j)}))$ and the set $\{1, ..., r\} \setminus K_{linear}$ of non-linear equations $(\sigma_{j_{1,k},k} \pi^{(j_{1,k})}, ..., \sigma_{j_{|Var(k)|,k},k} \pi^{(j_{|Var(k)|,k})}) \in H_k$ for all k belonging to it (where, once again, $Var(k) = \{j_{1,k}, ..., j_{|Var(k)|,k}\}$). It therefore implies the question whether such a system of equations is NP-compete.

To summarize, taking into account the fact that the proposed graph with "variables" and "equations" is just one of the simplest cases of the above-stated compacted graph construction (for instance, the introduced bipartite loop-graphs of "equations" could be further developed via adding initial edges inside their parts), we can say that we accordingly get an apparatus for formulating a variety of conjectures for both graph isomorphism and group theories whose correctness implies the equality GI = NP.

# A polynomial-time heuristic approach
# to the clique problem

The general types of techniques and notions that we applied when dealing with the graph isomorphism problem could also be applicable in resolving the much more important problems of determining a graph's clique number and finding its maximum clique, i.e. for the clique problem which is well-known to be NP-complete, unlike the graph isomorphism one.

Given a simple undirected graph G with n vertices whose adjacency matrix is A, we're going to build a sequence of real-valued n×n-matrices $X^{(q)} = X^{(q)}(G)$ (whose columns we'll consider as the coordinate-vectors of n points in $\mathbb{R}^n$, while the j-th point we'll associate with the vertex j of G) by the following recursive scheme:

$$X^{(0)} = I_n$$

Beginning with q = 1, at the step q of our recursive process, we define for j =1,…,n

(*)     $x_j^{(q+1)} = x_j^{(q)} + g \sum_{k \in \{1,...,n\} \setminus \{j\}} \frac{\varepsilon a_{jk}}{d^s(x_j^{(q)}, x_k^{(q)})} (x_k^{(q)} - x_j^{(q)})$

where $x_j^{(q)}$ is the j-th column of $X^{(q)}$, h, s, ε are the proposed algorithm's parameters (h, s, ε > 0), and for two n-vectors y, z **d**(y,z) denotes the Euclidean distance between them in $\mathbb{R}^n$.

The above recursive relation is, in fact, a computational circuit for numerically solving the autonomous system of differential equations

$$\dot{x}_j(t) = g \sum_{k \in \{1,...,n\} \setminus \{j\}} \frac{a_{jk}}{d^s(x_j(t), x_k(t))} (x_k(t) - x_j(t))$$

and the chief idea standing behind it is the **conjecture** that **after a certain period of time since the beginning of the initial point system's contraction under the influence of the introduced**

"gravitational forces" between pairs of points connected, as vertices, by G's edges the smallest distance will always appear between a pair of vertices belonging to a maximum clique of G.

A special interest this conjecture presents for the case of regular and semi-regular graphs, particularly for the graph generated by a CNF so that its vertex set is the CNF's set of literals and a pair of literals isn't to be connected by an edge if and only if either both of them belong to one disjunctive clause or they're the opposite powers of one variable (this graph's clique number equals the CNF's number of disjunctive clauses if and only if the CNF is satisfiable and is smaller otherwise, and the graph becomes close to regular upon bounding by constants the CNF's quantity of literals in a disjunctive clause and the number of times a variable can occur in literals, while it's regular when the two numbers are just constants, i.e. same for all the clauses and all the variables correspondingly). It's actually known that the clique problem is NP-complete for regular graphs, even when restricted to the case of graphs complimentary to cubic planar ones.

Hence such an algorithm is supposed to perform a certain polynomial (in n) number of steps (*) (considered as the algorithm's functional parameter), determine a pair of vertices (u,v) of the smallest distance, and construct the graph $G_{N_G(u,v)}$ received from G as the graph induced by the set $N_G(u,v)$ of common neighbors of u and v in G. (We additionally conjecture that, after a certain period of time since the gravitational contraction's beginning, any pair of vertices of the smallest distance is to belong to a maximum clique, although we may also suppose that graphs with no automorphism should get just one such pair at each moment). After that the whole process is to be repeated for $G_{N_G(u,v)}$ etc. until we get either an empty graph or a graph with one vertex.

An additional idea for enhancing the above approach to the clique problem may be introducing "repelling forces" (aka "anti-gravitational") between pairs of points that aren't connected, as vertices, in G. In such a case we'll receive the equation

$$\dot{x}_j(t) = g \sum_{k \in \{1,\ldots,n\} \setminus \{j\}} \frac{a_{jk}}{d^s\left(x_j(t), x_k(t)\right)} (x_k(t) - x_j(t)) -$$

$$- g_1 \sum_{k \in \{1,\ldots,n\} \setminus \{j\}} \frac{1 - a_{jk}}{d^{s_1}\left(x_j(t), x_k(t)\right)} (x_k(t) - x_j(t))$$

with the corresponding computational circuit for numerical solving

(**) $\quad x_j^{(q+1)} = x_j^{(q)} + g \sum_{k \in \{1,\ldots,n\} \setminus \{j\}} \frac{\varepsilon a_{jk}}{d^s(x_j^{(q)}, x_k^{(q)})} (x_k^{(q)} - x_j^{(q)}) -$

$$- g_1 \sum_{k \in \{1,\ldots,n\} \setminus \{j\}} \frac{\varepsilon(1 - a_{jk})}{d^{s_1}(x_j^{(q)}, x_k^{(q)})} (x_k^{(q)} - x_j^{(q)})$$

Let's call g and $g_1$ the *gravitation* and *anti-gravitation coefficients* correspondingly.

Besides, we can notice that the matrices $X^{(q)}(G)$ are, of course, invariant under any of G's automorphisms and, given two graphs $G_1$ and $G_2$, we accordingly may also use $X^{(q)}(G_1)$ and $X^{(q)}(G_2)$ for determining if they're isomorphic (via comparing $MSp(X^{(q)}(G_1))$ and $MSp(X^{(q)}(G_2))$ and eventually obtaining a pair of final algebras upon their entry spectrums' ceasing to grow), but, nevertheless, in such a case we're actually supposed to figure out whether it's not a partial case of the initial graph's adjacency matrix's modification via a series of meta-polynomial and entry spectrum replacement transformations, though, in any case, $X^{(q)}(G)$ is, of course, an isomorphism-commuting transformation of G's adjacency matrix.

The proposed algorithm for the clique problem had been tested (via computer modeling) on graphs received (as described above) from random CNF samples with several hundred Boolean variables whose maximal number of literals in a clause and maximal number of a variable's occurrence was 3 and showed correctness and polynomial-time performance in finding, in case of the CNF's satisfiability, a satisfying Boolean vector.

And, at last, it would be worth noting that any kind of approach to the clique problem can be further enhanced with a randomization parameter via embedding the given graph G with n vertices whose clique number we need to determine into the graph received from G through *bipartitely gluing* it with a random graph $G_1$ with $n_1$ vertices whose clique number we know, -- while we define the *bipartite gluing* of two graphs $G = (V, E)$, $G_1 = (V_1, E_1)$ for disjoint V, $V_1$ as the graph $G \oslash G_1 = (V \cup V_1, E \cup E_1 \cup K_{V,V_1})$ where $K_{V,V_1}$ is the complete bipartite graph on the parts V, $V_1$. The clique number of $G \oslash G_1$ obviously equals the sum of the clique numbers of G and $G_1$ and this gluing is $(k + n_1)$-regular when G is k-regular, $G_1$ is $k_1$-regular and $n + k_1 = n_1 + k$. Hence, in case if our algorithm works out on a certain sufficiently large fraction of q-regular graphs with m vertices for a certain set of values of the ratio q:m (containing NP-complete cases), we can also conjecture that such a random gluing may be quite capable of resolving, via the proposed "gravitational" algorithm, the most hard cases of instances with a sufficiently high (for being a polynomial-time randomized computational circuit) probability of success. However, the general direction of research regarding the above-stated gravitation contraction model may, of course, be rather related to attempting to understand the behavior of its differential equation's solution and even trying, on the basis of such understanding, to reduce the algorithm's polynomial-time complexity through modifying its vertex selection criterion for to take, at each global step, not just one pair, but a much bigger set of vertices as supposedly belonging to a maximum clique.

**References:**

Levin, Leonid (1986). "Average-case complete problems". SIAM J. Comput. 15 (1): 285–6. doi:10.1137/0215020.

Levin, Leonid (2014), "Computational Complexity of Functions" https://arxiv.org/abs/1411.3010

Levin, Leonid; Ramarathnam Venkatesan (1988), "Random instances of a graph coloring problem are hard",

STOC '88 Proceedings of the twentieth annual ACM symposium on Theory of computing Pages 217-222

Babai, László (1980), "On the complexity of canonical labeling of strongly regular graphs", SIAM Journal on Computing, 9 (1): 212–216, doi:10.1137/0209018, MR 0557839.

Babai, László; Codenotti, Paolo (2008), "Isomorphism of hypergraphs of low rank in moderately exponential time" (PDF), Proceedings of the 49th Annual IEEE Symposium on Foundations of Computer Science (FOCS 2008), IEEE Computer Society, pp. 667–676, doi:10.1109/FOCS.2008.80, ISBN 978-0-7695-3436-7.

Babai, László; Grigoryev, D. Yu.; Mount, David M. (1982), "Isomorphism of graphs with bounded eigenvalue multiplicity", Proceedings of the 14th Annual ACM Symposium on Theory of Computing, pp. 310–324, doi:10.1145/800070.802206, ISBN 0-89791-070-2.

Babai, László; Kantor, William; Luks, Eugene (1983), "Computational complexity and the classification of finite simple groups", Proceedings of the 24th Annual Symposium on Foundations of Computer Science (FOCS), pp. 162–171, doi:10.1109/SFCS.1983.10.

Babai, László; Luks, Eugene M. (1983), "Canonical labeling of graphs", Proceedings of the Fifteenth Annual ACM Symposium on Theory of Computing (STOC '83), pp. 171–183, doi:10.1145/800061.808746, ISBN 0-89791-099-0.

Babai, László (2015), Graph Isomorphism in Quasipolynomial Time, arXiv:1512.03547, Bibcode:2015arXiv151203547B * Baird, H. S.; Cho, Y. E. (1975), "An artwork design verification system", Proceedings of the 12th Design Automation Conference (DAC '75), Piscataway, NJ, USA: IEEE Press, pp. 414–420.

Blum, Manuel; Kannan, Sampath (1995), "Designing programs that check their work", Journal of the ACM, 42 (1): 269–291, doi:10.1145/200836.200880.

Bodlaender, Hans (1990), "Polynomial algorithms for graph isomorphism and chromatic index on partial k-trees", Journal of Algorithms, 11 (4): 631–643, doi:10.1016/0196-6774(90)90013-5, MR 1079454. * Booth, Kellogg S.; Colbourn, C. J. (1977), Problems polynomially equivalent to graph isomorphism, Technical Report, CS-77-04, Computer Science Department, University of Waterloo.

Booth, Kellogg S.; Lueker, George S. (1979), "A linear time algorithm for deciding interval graph isomorphism", Journal of the ACM, 26 (2): 183–195, doi:10.1145/322123.322125, MR 0528025.

* Boucher, C.; Loker, D. (2006), Graph isomorphism completeness for perfect graphs and subclasses of perfect graphs (PDF), Technical Report, CS-2006-32, Computer Science Department, University of Waterloo.

Abello, J.; Pardalos, P. M.; Resende, M. G. C. (1999), "On maximum clique problems in very large graphs" (PDF), in Abello, J.; Vitter, J., External Memory Algorithms, DIMACS Series on Discrete Mathematics and Theoretical Computer Science, 50, American Mathematical Society, pp. 119–130, ISBN 0-8218-1184-3.

Amano, Kazuyuki; Maruoka, Akira (2005), "A superpolynomial lower bound for a circuit computing the clique function with at most (1/6)log log N negation gates", SIAM Journal on Computing, 35 (1): 201–216, doi:10.1137/S0097539701396959, MR 2178806.

Arora, Sanjeev; Lund, Carsten; Motwani, Rajeev; Sudan, Madhu; Szegedy, Mario (1998), "Proof verification and the hardness of approximation problems", Journal of the ACM, 45 (3): 501–555, doi:10.1145/278298.278306, ECCC TR98-008. Originally presented at the 1992 Symposium on Foundations of Computer Science, doi:10.1109/SFCS.1992.267823.

Arora, S.; Safra, S. (1998), "Probabilistic checking of proofs: A new characterization of NP", Journal of the ACM, 45 (1): 70–122, doi:10.1145/273865.273901. Originally presented at the 1992 Symposium on Foundations of Computer Science, doi:10.1109/SFCS.1992.267824.

Balas, E.; Yu, C. S. (1986), "Finding a maximum clique in an arbitrary graph", SIAM Journal on Computing, 15 (4): 1054–1068, doi:10.1137/0215075.

Barrow, H.; Burstall, R. (1976), "Subgraph isomorphism, matching relational structures and maximal cliques", Information Processing Letters, 4 (4): 83–84, doi:10.1016/0020-0190(76)90049-1.

Battiti, R.; Protasi, M. (2001), "Reactive local search for the maximum clique problem", Algorithmica, 29 (4): 610–637, doi:10.1007/s004530010074. * Bollobás, Béla (1976), "Complete subgraphs are elusive", Journal of Combinatorial Theory, Series B, 21 (1): 1–7, doi:10.1016/0095-8956(76)90021-6, ISSN 0095-8956.

Boppana, R.; Halldórsson, M. M. (1992), "Approximating maximum independent sets by excluding subgraphs", BIT Numerical Mathematics, 32 (2): 180–196, doi:10.1007/BF01994876.

Cazals, F.; Karande, C. (2008), "A note on the problem of reporting maximal cliques" (PDF), Theoretical Computer Science, 407 (1): 564–568, doi:10.1016/j.tcs.2008.